\documentclass[reqno,12pt]{amsart} 

\usepackage{epsf}\usepackage{amssymb} \usepackage{cite}

\numberwithin{equation}{section}
\newcommand{\beq}{\begin{equation}}
\newcommand{\ee}{\end{equation}}
\newcommand{\bea}{\begin{eqnarray}}
\newcommand{\eea}{\end{eqnarray}}

\usepackage{hyperref}
\def\stackreb#1#2{\ \mathrel{\mathop{#1}\limits_{#2}}}

\newcommand{\CC}{\mathbb C}
\newcommand{\R}{\mathbb R}
\newcommand{\Z}{\mathbb Z}

\newcommand{\la}{\langle}
\newcommand{\ra}{\rangle}
\renewcommand{\beq}{\begin{equation}}
\renewcommand{\ee}{\end{equation}}
\newcommand{\ba}{\begin{eqnarray}}
\newcommand{\ea}{\end{eqnarray}}
\newcommand{\baa}{\begin{eqnarray*}}
\newcommand{\eaa}{\end{eqnarray*}}
\newcommand{\bb}{\begin{thebibliography}}
\newcommand{\eb}{\end{thebibliography}}

\def\stackreb#1#2{\ \mathrel{\mathop{#1}\limits_{#2}}}

\newcommand{\tb}   {{\ifmmode \tan\beta     \else $\tan\beta$      \fi}}
\def\bea{\begin{eqnarray}}
\def\eea{\end{eqnarray}}

\begin{document}

\title[Self-similar potentials]
{\bf Self-similar potentials in quantum mechanics
 \\[1mm]
 and coherent states}

\author{V. P. Spiridonov}

\date{}

\vspace{-2em}

\makeatletter
\renewcommand{\@makefnmark}{}
\makeatother
\footnotetext{
Based on the talk given at the seminar ``In the search of beauty: from condensed matter
to integrable systems'', dedicated to V. B. Priezzhev, 10 September 2019, BLTP JINR;
http://thproxy.jinr.ru/video/seminars/2019-09-10/mp4/ \\   \indent $\;\;$
This work is partially supported by the Laboratory of Mirror Symmetry NRU HSE, RF government grant,
ag. no. 14.641.31.0001.}

\address{Laboratory of theoretical physics, JINR, Dubna;
National Research University Higher School of Echonomics, Moscow.}

\begin{abstract}
A brief description of the relations between the factorization method in quantum mechanics,
self-similar potentials, integrable systems and the theory of special functions is given.
New coherent states of the harmonic oscillator related to the Fourier trans\-for\-ma\-tion are constructed.
\end{abstract}

\maketitle

\tableofcontents

\vspace*{1em}

\begin{flushright}
\em To the blessed memory of V. B. Priezzhev
\end{flushright}


\section{Evolution of Schr\"odinger operators in discrete time}

Let us consider the problem of solving stationary Schr\"odinger equation in one-dimensional space
\beq
L\psi(x)=\lambda\psi(x), \qquad \lambda, x\in\mathbb{R},
\label{Sch}\ee
where the Hamiltonian $L$ has the form (in appropriate normalization of the coordinate
$x$ and the energy $\lambda$)
$$
L=-\partial_x^2 +u(x), \qquad \partial_x:=\frac{d}{dx},\quad u(x)\in C^\infty.
$$
Here, for concreteness, we have indicated that the potential $u(x)$ is a real infinitely
diffe\-ren\-ti\-able function. In fact, this is not obligatory, the potential may be described
by a discontinuous function or even by a tempered distribution. The main physical problem
consists in determination of the energy spectrum, i.e. the values of $\lambda$ for which
the wave functions (or eigenfunctions) $\psi(x)$ are either square integrable, i.e. lie
in the Hilbert space $\mathrm{L}^2(\mathbb{R})$, or they do not grow at the infinity faster than
a power function.

From the purely mathematical viewpoint it is interesting to characterize the class of
potentials $u(x)$ for which the wave functions can be constructed in the closed form
for all $\lambda$. This means that for $\psi(x)$ there should exist an expression either
in the form of some series with simple coefficients, or in the form of definite integrals
with explicit integrand functions. Here, ``simple'' and ``explicit'' mean that the divisor
points of these coefficients or integrands are known and they are described by
sequences of numbers expressed through elementary functions (for example, by arithmetic
or geometric progressions).   In the framework of this problem it is convenient to use the power
of complex analysis and analyze equation \eqref{Sch} in the general case, when $x, \lambda\in\CC$
and consider $u(x)$ as a complex analytical function at least in some compact domain of $x$.

A constructive approach to this problem consists in the
investigation of symmetries of the Schr\"odinger equation not for a concrete potential, but
for the space of all potentials at once. For this let us consider an evolution of the Schr\"odinger
operators in some artificial discrete time $j\in\Z$ and construct an infinite chain of Hamiltonians
\beq
L_j=-\partial_x^2 +u_j(x), \qquad L_j\psi^{(j)}(x)=\lambda \psi^{(j)}(x), \quad j\in \mathbb{Z},
\label{DTL}\ee
whose wave functions are related to each other by action of the differential operators
of the first order
\beq
\psi^{(j+1)}(x)=A_j\psi^{(j)}(x), \qquad A_j=\partial_x+f_j(x).
\label{DTE}\ee
It is evident that for arbitrary $u_j(x)$ and $f_j(x)$ the pair of equations \eqref{DTL} and \eqref{DTE}
(this pair was introduced by Infeld \cite{Inf}, but in the theory of integrable systems such
pairs are called ``the Lax pairs'') contradict to each other. The conditions under which there
emerges a compatible system have the form
\beq
L_{j+1}\psi^{(j+1)}=L_{j+1}\big[A_j\psi^{(j)}\big]=\lambda \big[ A_j\psi^{(j)}\big] =A_jL_j\psi^{(j)},
\label{comp}\ee
which lead to the intertwining relations in the operator form
\beq
L_{j+1}A_j=A_jL_j.
\label{inter1}\ee
Here, the left-hand and right-hand side expressions are differential operators of the third order.
Equating the functional coefficients in front of different powers of $\partial_x$ we
obtain the equations
$$
u_{j+1}(x)=u_j(x)+2f'_j(x), \qquad u_j'(x)=2f_j(x)f_j'(x) -f_j''(x),
$$
where the primes mean the derivatives with respect to $x$. After solving them we find the
explicit connection between $u_j(x)$ and $f_j(x)$:
\beq
u_j(x)=f_j^2(x)-f_j'(x)+\lambda_j, \qquad u_{j+1}(x)=f_j^2(x)+f_j'(x)+\lambda_j,
\label{ujfj}\ee
where $\lambda_j$ are integration constants.

It is easy to see that, as a result of resolving of the compatibility conditions, we came to
a natural factorization of the Hamiltonians
\beq
L_j=A_j^+A_j+\lambda_j, \qquad L_{j+1}=A_jA_j^+ +\lambda_j,
\label{fact}\ee
where the operators $A_j^+$ are formal Hermitian conjugates of $A_j$,
\beq
A_j^+=-\partial_x+f_j(x),\qquad  A_j^+ L_{j+1}=L_j A_j^+.
\label{Aconj}\ee
If $\lambda_j\in\R$ and $f_j(x)$ do not contain singularities, then $A_j^\dag=A_j^+$
and $ L_j^\dag=L_j$.

After replacing $j$ by $j+1$ in the first expression in \eqref{ujfj} and equating it to the second one
we obtain the Infeld factorization chain \cite{Inf}
\beq
A_{j+1}^+A_{j+1}+\lambda_{j+1}=A_jA_j^+ + \lambda_j,
\label{factchain}\ee
or
\beq
 f_{j+1}^2(x)-f_{j+1}'(x)+\lambda_{j+1}=f_j^2(x)+f_j'(x)+\lambda_j,
\label{chain}\ee
defining the basic equation of the factorization method in quantum mechanics,
which was formulated first by Schr\"odinger on concrete examples in \cite{Sc1,Sc2}.

Historically, the trans\-for\-ma\-tions mapping solutions of certain linear differential equa\-tions
to other linear differential equations, different from the initial ones, were considered
long before creation of quantum mechanics. In the integrable systems literature
they are conventially called the Darbous trans\-for\-ma\-tions. Let us pass to
description of the essence of the factorization method.

\section{The factorization method}

Let us consider how factorization of Hamiltonians to a product of differential
operators of the first order helps to determine the discrete spectrum of a  number
of simple Schr\"odinger operators in one-dimensional quantum mechanics.
A detailed description of this method was given in the well known old survey \cite{IH}.

The key example is the one-dimensional harmonic oscillator -- the simplest quantum mechanical system
carrying a universal character.  The Hamiltonian of this system in the appropriate  normalization
of the canonical coordinates and energy units ($\hbar =\omega=m=1$) has the form
\beq
L=\tfrac{1}{2}\left(p^2+x^2\right), \qquad [x,p]=xp-px=\textup{i}, \qquad \textup{i}^2=-1.
\label{harmHam}\ee

The eigenvalue problem for a Hamiltonian, $L\psi=\lambda\psi$, determines admissible energy values
of the system $\lambda$. If the eigenfunctions $\psi$ are normalizable vectors in the Hilbert
space, then they describe the discrete (or point) spectrum. For an absolutely continuous spectrum
$\psi$ lie in the rigged Hilbert space containing tempered distributions (we do not touch
the case of the singular continuous spectrum).   One can pass to the coordinate representation
for which $x\in\mathbb{R}$, $p=-\textup{i}\partial_x $ and $\psi=\psi(x)$. Then
$L\psi=\lambda\psi$ transforms to the Schr\"odinger equation having the form of the differential
equation of second order, general solution of which for the operator
\eqref{harmHam} is expressed in terms of the confluent hypergeometric function.
Normalizable cases $\psi(x)\in L^2(\mathbb{R})$ are easily singled out and lead to the Hermite polynomials.

However, this way of solving the eigenvalue problem looks overcomplicated. It can be solved
by a purely algebraic method. Let us factorize the Hamiltonian \eqref{harmHam}
\beq
L=a^+a + \tfrac{1}{2},\quad a^+= \frac{-\textup{i}p+x}{\sqrt{2}}, \quad a= \frac{\textup{i}p+x}{\sqrt{2}}.
\label{harmFact}\ee
The operators $L, a, a^+$ form a simple Lie algebra called the Heisenberg algebra,
 which follows from the canonical commutation relations \eqref{harmHam}:
\beq
[a, a^+]=1, \qquad [L,a]=-a,\qquad [L,a^+]=a^+.
\label{Heis}\ee
Let us denote $|0\ra$ the  ground state of the system, or the vacuum, defined as the zero mode
of the operator $a$, $a|0\ra=0$. We shall assume that this state is normalized to unity
$\la 0|0\ra=1$. Then we find the complete systems of eigenfunctions of the Hamiltonian simply
by a sequential action of the raising operator $a^+$:
\beq
L|n\ra=\lambda_n|n\ra,\quad \lambda_n=n+\tfrac{1}{2},\quad |n\ra=\frac{1}{\sqrt{n!}}(a^+)^n\,|0\ra,\qquad \la n|m\ra=\delta_{nm}.
\label{harmEnergy}\ee
Actions of the raising $a^+$ and lowering $a$ operators have the following explicit form
\beq
a^+|n\ra=\sqrt{n+1}\,|n+1\ra,\qquad a|n\ra=\sqrt{n}\,|n-1\ra.
\label{ladder}\ee

In the coordinate representation we have $a=\frac{\partial_x +x}{\sqrt{2}},
\, a^+=\frac{-\partial_x +x}{\sqrt{2}}$ and, correspondingly,
\beq
\psi_n(x):=\la x |n\ra=\frac{H_n(x)}{\sqrt{2^nn!\sqrt{\pi}}}e^{-x^2/2}, \qquad
H_n(x)=(-1)^n e^{-x^2}\frac{d^n}{dx^n} e^{x^2},
\label{harmEigen}\ee
where $H_n(x)$ are the Hermite orthogonal polynomials.

In this example the same operator $a$ tranfers from the $n$-th state
to the state of the number $n-1$. In the general case of the factorization method
the form of the lowering (or raising) operator depends on the level number. Let us consider
zero modes of the generalized lowering operators $A_j$ \eqref{DTE},
\beq
A_j\psi^{(j)}_{vac}=0 \qquad \Rightarrow \quad \psi^{(j)}_{vac}(x)=e^{-\int^x_{x_0} f_j(y)dy}.
\label{vac}\ee
Suppose that all sequential zero modes starting from  $j=0$ and to
some number $N>0$ are normalizable functions, $\psi^{(j)}_{vac}\in {\rm L}^2(\mathbb{R})$,
i.e. they describe discrete spectrum states. Then
\[
L_j\psi^{(j)}_{vac}=\lambda_j\psi^{(j)}_{vac}
\]
and the integration constants $\lambda_j$ appear to be equal to the ground state energies of the
Hamiltonians $L_j$. The generalized raising operators  $A_j^+$  allow one to build other exciting states
of the system. Let us consider the state $\psi_1^{(j)}:= A_j^+\psi^{(j+1)}_{vac}$. For this state
\[
L_j\psi_1^{(j)}=A_j^+L_{j+1}\psi^{(j+1)}_{vac}=\lambda_{j+1}A_j^+\psi^{(j+1)}_{vac}=\lambda_{j+1}\psi_1^{(j)},
\]
i.e. $\lambda_{j+1}$ is an eigenvalue of the operator $L_j$ (the second level of the discrete spectrum).

Consider the evolution operators describing development in the discrete time to $n>0$ steps,
\beq
M_j^{(n)}:=A_{j+n-1}\cdots A_{j+1}A_j, \qquad
M_j^{(n)+}:=A_j^+A_{j+1}^+\cdots A_{j+n-1}^+.
\label{nstep}\ee
For them one has the following intertwining relations following from
equalities \eqref{inter1} and \eqref{Aconj},
\beq
L_{j+n}M_j^{(n)} = M_j^{(n)}L_j, \qquad M_j^{(n)+}L_{j+n}=L_jM^{(n)+}_j.
\label{interN}\ee
It is easy to derive the operator factorization relations
\beq
M_j^{(n)+}M_j^{(n)}=\prod_{k=0}^{n-1}(L_j-\lambda_{j+k}), \quad
M_j^{(n)}M_j^{(n)+}=\prod_{k=0}^{n-1}(L_{j+n}-\lambda_{j+k}).
\label{poly}\ee
The  forward and backward in discrete time compatible evolution equations look as follows
\beq
\psi^{(j+n)}=M_j^{(n)}\psi^{(j)}, \quad
\psi^{(j)}=\prod_{k=0}^{n-1}(\lambda-\lambda_{j+k})^{-1}\; M_j^{(n)+}\psi^{(j+n)}.
\label{evolution}\ee
Also, from the algebraic intertwining relations \eqref{interN} it naturally follows
that all wave functions $\psi^{(j+k)}$ are eigenfunctions of the initial Hamiltonian $L_j$
after an appropriate number of steps $j\to j-1$,
$$
L_j\psi_n^{(j)}=\lambda_{j+n}\psi_n^{(j)}, \qquad \psi_n^{(j)}:=M_j^{(n)+}\psi^{(j+n)}_{vac}.
$$
That is $\lambda_j,\, \lambda_{j+1}, \ldots$ describe the discrete spectrum of the Hailtonian $L_j$.
For this statement to be true one has to have normalizable functions, $\psi_n^{(j)}\in \mathrm{L}^2(\mathbb{R})$, $n=0, 1, \ldots.$ At the same time it is not difficult to check that
this is the full spectrum, i.e. there are no any missing eigenvalues. A check of the normalizability
condition requires knowledge of asymptotics of the functions $f_j(x)$ and absence of
singularities on the finite real axis.

If one is interested in general solutions of the Schr\"odinger equations
then we draw attention to the fact that starting from a general solution of an initial equation
$L_0\psi^{(0)}=\lambda\psi^{(0)}$,
the present formalism allows one to build new solvable equations $L_m\psi^{(m)}=\lambda\psi^{(m)}$, $m\in\Z$.
During this procedure the spectrum of the operators $L_m,\, m>0,$ coincide with the spectrum of
$L_0$ with the first $m$ eigenvalues deleted. Evidently, starting from the Hamiltonian with
a given spectrum $L$ one can construct new Hamiltonians not with deleted but, vice versa, with
inserted new levels -- for this it is sufficient to reverse the refactorization
procedure. If one violates the condition that $\psi_{vac}^{(j)}$ are normalizable functions,
then it is possible to build the isospectral Hamiltonians with different potentials. Moreover,
one can add or remove an arbitrary number of discrete levels not only with the lowest
energy values, but, under certain conditions, in any part of the spectrum \cite{Krein}.
Therefore, the Infeld factorization chain can be called the chain of spectral trans\-for\-ma\-tions
of the Schr\"odinger equation.

Evidently, in the transition from $L_0$ to $L_m$ a part of the solutions corresponding to
particular values of $\lambda$ is ``lost''. Namely, for $\lambda=\lambda_j$,  $j=0, 1,\ldots, m-1$,
only one of the independent solutions of the equation $L_m\psi^{(m)}=\lambda\psi^{(m)},\, m>0,$
can be represented in the form $M_0^{(m)}\psi^{(0)}$, because the functions
$\psi^{(0)}_{vac}, A_0^+\psi^{(1)}_{vac},\ldots, A_0^+\cdots A_{m-2}^+\psi^{(m-1)}_{vac}$
describe zero modes of the evolution operator $M_0^{(m)}$. However, it is easy to restore
an explicit form of these missing solutions. Let $\psi_1, \, \psi_2$ be two independent
solutions of the equation $L\psi=\lambda\psi$ for some fixed $\lambda$, i.e.
$W(\psi_1, \psi_2)=\psi_1 \psi_2'-\psi_1' \psi_2\neq 0$. If one knows only one solution
$\psi_1(x)$, then the second one is restored from the first order equation
determined by the Wronskian condition $W=const$ by the formula.
$$
\psi_2(x)=\psi_1(x)\int^x_{x_0}\frac{dy}{\psi_1^2(y)}.
$$

\section{Self-similar potentials}

Basics of the factorization method were laid down by Schr\"odinger, however, a universal
approach to it was developed by Infeld \cite{Inf} by the formulation of the factorization chain
\eqref{chain}, which can be conveniently rewritten in the form
\beq
f_j'(x)+f_{j+1}'(x)+f_j^2(x)-f_{j+1}^2(x)=\mu_j,\quad \mu_j:= \lambda_{j+1}-\lambda_j.
\label{FC}\ee
The key moment consisted in the suggestion to consider in this chain the symbol
$j$ not as a discrete variable, but as a continuous parameter and to treat  \eqref{FC}
as a diffential-difference equation over two (in general complex) variables $x$ and $j$.

Correspondingly, one can search for solutions of the chain \eqref{FC} in the form of
a finite Laurent series in the variable $j$, $f_j(x)=\sum_{n=-N}^M a_n(x)j^n$,
for some integers $N$ and $M$. It appeared that solutions of such form for fininte $N$ and $M$
exist only if $N=M=1$, that is for the following ansatz of the generalized separation of variables
\beq
f_j(x)=\frac{a(x)}{j} +b(x)+ c(x)j.
\label{SepVar}\ee
After substitution of this expression in \eqref{FC} there emerge some equations
on the coefficients $a(x),$ $b(x),$ $c(x)$ and eigenvalues $\lambda_j$.
These equations are solved in terms of elementary functions and result in the
spectra of the form $\lambda_j\propto\, j,\, j^2, \alpha j^2+\beta/j^2,$
corresponding to the harmonic oscillator, the P\"oschl-Teller potential, the
Coulomb problem, and so on. The detailed description of the emerging
systems is given by Infeld and Hull in \cite{IH} and we shall not repeat it here.

Let us describe another class of solutions of \eqref{FC}, which we shall be calling self-similar
solutions. Let us consider the simplest symmetries of the Infeld chain. First, because there is
no explicit dependence on the discrete time variable $j$, one has the translational invariance:
the shift of the discrete time  $j\to j+N,\, N\in\mathbb{Z}$, maps solutions of this
chain to its other solutions. Another simple symmetry deals with the coordinate variable
and is related to the affine trans\-for\-ma\-tions $x\to qx+h$.
Namely, the trans\-for\-ma\-tion
\[
x\to qx+h, \qquad f_j(x)\to qf_j(qx+h),\quad \mu_j\to q^2\mu_j,
\]
maps given solutions of the chain   \eqref{FC} to its other solutions.

The general highly nontrivial self-similar solutions of the chain of equations
\eqref{FC} are defined as special solutions which are invariant under a combination
of these quite trivial symmetries \cite{spi:nasa} (see also \cite{spi:universal}):
\beq
f_{j+N}(x)=qf_j(qx+h), \qquad \mu_{j+N}= q^2\mu_j.
\label{reduction}\ee
Using a freedom in the definition of eigenvalues -- the possibility to shift them
by an arbitrary constant  $\lambda\to \lambda+const$, the shape of the discrete spectrum
following from the second equality in  \eqref{reduction} can be represented in
the following form
\beq
\lambda_{kN+m}=\lambda_m q^{2k}, \quad  m=0,1,\ldots,N-1,\; k=0,1, \ldots,
\label{spectrum}\ee
that is the expected discrete spectrum consists of a superposition of  $N$
geometric progressions with the increment $q^2$, under the condition
that $\lambda_j,\, q\in\mathbb{R}$ and $\lambda_0<\lambda_1<\ldots<\lambda_{N-1},$
and that the potentials $u_j(x)=f_j^2(x)-f'_j(x)+\lambda_j$ do not
contain singularities for $x\in\mathbb{R}$. In principle, the true number of
geometric progressions in the spectrum may be smaller than $N$, it depends on the
concrete choice of the potential parameters.

Because the discrete spectrum  $\lambda_n$ cannot grow faster than $n^2$ for $n\to\infty$,
for real physical situation it is necessary to impose the constraint  $0 < q^2 < 1$.
In this case the spectral parameter value $\lambda=0$ is the accumulation point of the discrete
spectrum, and for $\lambda>0$ there must be the continuous spectrum. All this is a qualitative
consideration requiring a rigorous proof by investigating the analytical structure of
the functions $f_j(x)$, including computation of their asymptotics for $x\to\pm \infty$.

\section{Quantum algebras (operator self-similarity)}

One can reformulate the described functional self-similarity using the operator language.
Let us impose an abstract constraint
\beq
L_{j+N}=q^2UL_jU^{-1}+\mu,
\label{OperRed}\ee
where $U$ is some invertible operator, and $q^2$ and $\mu$ are some constants.
If one shifts $L_j\to L_j+\frac{\mu}{1-q^2}$, then for $q^2\neq1$ the constant $\mu$
can be removed from the consideration. Let us denote $L:=L_0,\, B:=U^{-1}M_0^{(N)},\, B^+:=M_0^{(N)+}U.$
Then from the relations  \eqref{interN} and \eqref{poly} it follows
\begin{eqnarray} \nonumber  &&
LB^+=q^2B^+L +\mu B^+,\qquad BL=q^2LB+\mu B,\quad
\\ &&
B^+B=\prod_{j=0}^{N-1}(L-\lambda_j),\qquad
BB^+=\prod_{j=0}^{N-1}(q^2L+\mu-\lambda_j).
\label{algebra}\end{eqnarray}

For $N=1$ this is nothing else than the $q$-harmonic oscillator algebra
\beq
AA^+ - q^2A^+A=1,\qquad  A=\frac{B}{\sqrt{\mu+\lambda_0(q^2-1)}},\quad
A^{+}:=\frac{B^{+}}{\sqrt{\mu+ \lambda_0(q^2-1)}}.
\label{qosc}\ee
As far as the author knows, this algebra was investigated in detail for the first time
in the discussion of a sufficiently exotic subject in the paper entitiled ``Parastochastics''
 \cite{FB}. Sporadically it appeared in some other fields \cite{Coo,AC}.
However, it became widely popular only after the paper  \cite{Mac}, which appeared on the
background of a general interest to the quantum groups. The algebraic scheme of derivation
of quantum algebras presented above on the basis of the factorization method was
constructed in \cite{S0,CAP,spi:symmetries}.

For the algebra \eqref{qosc} it is natural to define the Hamiltonian in the form
\beq
H =A^+A-\frac{1}{1-q^2}=\frac{ L-\frac{\mu}{1-q^2}}{\mu+ \lambda_0(q^2-1)},
\quad L=B^+B+\lambda_0,
\label{qHam}\ee
which satisfies the relations $HA^+=q^2A^+H$ and $AH=q^2HA$. In the general situation one can consider
also the non-unitary case, when  $A^+\neq A^\dag$. In particular, this happens for
arbitrary complex $q\in\CC$, including into itself the most interesting
arithmetic systems when $q$ is a primitive root of unity, $q^m=1$, $m\in\Z_{\geq 0}$.

If one takes the formal vacuum state $A|0\ra=0$, $\la 0|0\ra=1$, the $q$-analogues of the bound
states of the harmonic oscillator are the following formal Hilbert space states
\beq
|n\ra=\frac{(A^+)^n}{\sqrt{[n]_{q^2}!}}|0\ra,
\quad [n]_{q^2}=\frac{1-q^{2n}}{1-q^2}, \quad [n]_{q^2}!=[1]_{q^2}[2]_{q^2} \cdots [n]_{q^2},
\label{qstates}\ee
with the spectrum
\beq
H|n\ra=\lambda_n|n\ra, \qquad \lambda_n=\frac{q^{2n}}{q^2-1},  \quad \la n|m\ra=\delta_{nm},
\label{qspectrum}\ee
which is derived by purely algebraic means.
In this situation both options are admissible, either $0<q^2<1$ or  $q^2>1$.
However, concrete realizations of the operator algebra impose certain restrictions
on the actual spectrum of states of the Hamiltonian  $H$. In particular, for the
ordinary Schr\"odinger equation only the choice $0<q^2<1$ is admissible.

\section{Special cases}

Let us consider special cases of self-similar potentials for which either the
complete solution of the spectral problem is known, or there exists at least some
useful information about them.

Let
\beq
q=1,\quad \mu=0, \quad U=1, \quad [L,B]=[L,B^+]=0.
\label{perclos}\ee
These restrictions, demanding existence of a nontrivial integral of motion in
the form of a differential operator of $N$-th order, lead to (for odd $N$
and, under certain conditions, for even $N$) the finite-gap potentials which are described by
the hyperelliptic functions. Setting  $N=2n+1$ or $N=2n$, one has
\beq
u_j(x)=-2\partial_x^2 \log \Theta(x), \qquad \Theta(x)
=\sum_{m_1,\ldots, m_n\in\Z} e^{2\pi\textup{i}( \sum_{k,\ell=1}^n A_{k\ell}m_km_\ell+\sum_{k=1}^nB_k(x)m_k)}.
\label{finitegap}\ee
The matrix of quasiperiods $A_{ij}$ of the Riemann theta function of genus $n$ and
variables $B_k(x)$ linearly depending on $x$ determine the potential shape and the Hamiltonian
spectrum \cite{DMN}.

Consideration of these potentials on the basis of a periodic reduction of the Infeld chain
and analysis of the corresponding system of differential equation as a completely
integrable system is given in  the papers \cite{Weiss} for $\mu_j=0$ and in \cite{VS}
for arbitrary $\lambda_{j+N}=\lambda_j$. The potential \eqref{finitegap} is called
the ``finite-gap'' because it is assumed that one solves the spectral problem
with periodic or quasiperiodic potentials, when the spectrum contains a finite
number of admitted (or forbidden) zones of the continuous spectrum. The spectral
problem of a different type related to boundary conditions at the
singular points of the potential, for which the spectrum is purely discrete,
was discussed in the work \cite{sko-spi:spectra}.

Already the case when $U$ is chosen in the form of the parity operator $U=P$, $Pg(x)=g(-x)$,
which corresponds to the dilation operator $Ug(x)=g(qx)$ for $q=-1$,
causes certain difficulties in the description of the explicit form of potentials.
Evidently, in this case $B^2$ and $( B^{+} )^2 $ will be differential operators of the order $2N$
commuting with the Hamiltonian.
If one chooses some even finite-gap potentials, corresponding to antisymmetric
functions $f_j(-x)=-f_j(x)$, then this yields a solution of the problem, but
a question on the existence of other solutions remains open.

Let
\beq
q=1,\quad \mu\neq 0, \quad U=1,\quad [L,B^+]=\mu B^+, \quad [L,B]=-\mu B.
\label{painleve}\ee
The corresponding Hamiltonians generalize the harmonic oscillator problem,
which emerges for $N=1$. The fact that the commutator algebras of the type \eqref{painleve}
are related to the Painlev\'e transcendents was established in the paper \cite{Fl}.
In the situation described by us for the periodic closure with $N=3$ the potential is
expressed in terms of the function Painlev\'e-IV  \cite{Bu}, and for $N=4$ one needs
already the Painlev\'e-V function with some small constraint \cite{VED}.
Formally the spectrum of bound states of the corresponding Hamiltonians consists of the
superposition of $N$ arithmetic progressions. However, this statement depends on
the boundary conditions, and already for $N=2$ it is not true because of the
singularities of the emerging potential. Such spectrum corresponds also to the choice
$q=-1$, when $U$ is the parity operator, $Uf(x)=f(-x)$, but now one has to be able to single out
solutions with particular parity properties  among the Painlev\'e type functions emerging from
the closure periods $2N$ \cite{spi:universal}.

Let $h\in\R$ and
\beq
q=1, \quad U=e^{h\partial_x}, \quad Uf(x)=f(x+h), \quad U^\dag=U^{-1}=e^{-h\partial_x}.
\label{hpainleve}\ee
The general solution is not known in any explicit form.
It is evident that there emerges some translational  $h$-deformation of the Painlev\'e
transcendents and their generalizations. For instance, at $N=1$ we have the
following equation for the function $f(x):=f_0(x)$,
\beq
f'(x)+f'(x+h)+f^2(x)-f^2(x+h)=\mu.
\label{trans}\ee
For $\mu=0$ we obtain the addition formula for the elliptic Weierstrass function $\wp(x)$:
$$
f(x)=-\frac{1}{2}\frac{\wp'(x-x_0) - \wp'(h)}{\wp(x-x_0)-\wp(h)}.
$$
Note that this function is not defined in the limit $h\to 0$, which demonstrates some
peculiarities  of concrete functional realizations of operator algebras which do not
pass to each other as the naively taken limits suggest. For the case  $\mu\neq0$
it was shown in the paper  \cite{tovbis} that for appropriate boundary conditions
$f(x)$ is an analytical function meromorphic on the whole complex plane,
which is a rather rare phenomenon for nonlinear differential-difference equations.

For arbitrary $N$ and $\mu=0$ there emerges a curious deformation of the general hyperelliptic
functions, which is still not properly investigated because now the integral of motion
is not the differential operator, but a differential-difference operator.

Let $U=q^{x\partial_x}$ and $q^{m}=1$ (a non-unitray case). Then
\beq
U^{m}=1, \qquad [L,B^{m}]=[L,(B^{+})^{m}]=0.
\label{root1}\ee
As a result, we obtain a special class of the finite-gap potentials (because $B^{m},\, (B^{+})^m$
are differetial operators), which are characterized by the presence of
rotation symmetry axes of the $m$-th order \cite{sko-spi:self}.

Let $q\in\CC^\times$ takes arbitrary values, then there exists a trivial particular solution
of the form $f_j(x)=const$, namely
\beq
U^{\pm1}=q^{\pm x\partial_x}, \quad U^{\pm1}f(x)=f(q^{\pm1}x),\quad f_j^2 = -\lambda_j, \quad u_j(x)=0.
\quad  L_j=-\partial_x^2,
\label{free1}\ee
As a result we obtained a trivial free quantum particle for which there is
a highly nontrivial polynomial algebra of symmetries, which for $N=1$
coincides with the algebra of $q$-harmonic oscillator.

Another case of simple explicitly known potentials corresponds to the crystal
basis of quantum algebras $q=0$. For nonsingular self-similar potentials this
corresponds to the case $f_N(x)=0$, which does not mean that the functions
$f_j(x)$, $j=0,\ldots, N-1$, are trivial. In fact, in this case the functions
$f_{N-n}(x)$ describe the well known $n$-soliton potentials of the Korteweg-de Vries equation
(KdV) considered below.

General case of the self-similar potentials
$$
U^{\pm1}=q^{\pm x\partial_x}, \quad U^{\pm1}f(x)=f(q^{\pm1}x), \quad q\neq 0,\quad q^m\neq 1,
$$
corresponds to the (continuous) $q$-deformed Painlev\'e functions (for $N=3,4$) and
their generalization to the transcendents of higher order. For $N=1,\, h=0$ the conditions \eqref{reduction}
are equivalent to the reduction $f_j(x)=q^jf(q^jx),\, \lambda_j=\lambda_0q^{2j}$,
which was constructed first in \cite{Sh}. In this situation practically there is no any information
on the global analytical properties of the functions $f_j(x)$. For $N=1$ one can show that
under certain boundary conditions the function $f(x)$ is a meromorphic function on
the whole complex plane, analogously to solutions of the equation \eqref{trans}.
Note that in general the operator $U$ is not unitary. Only the operator $Uf(x)=|q|^{1/2} f(qx)$
will be unitary for $0<q<1$ and $-1<q<0$.

\section{Coherent states}

In the work \cite{Sch1} Schr\"odinger has constructed an overcomplete
system of states of the harmonic oscillator called the coherent states, which
appear to be normalized eignenstates of the lowering operator
\beq
A|\alpha\ra=\alpha|\alpha\ra, \quad \alpha\in\CC, \qquad \la\alpha|\alpha\ra=1.
\label{coherent}\ee
In the algebraic representation
\beq
|\alpha\ra=e^{\alpha a^\dagger-\alpha^*a}|0\rangle=e^{-\tfrac{1}{2}|\alpha|^2}\sum_{n=1}^\infty\frac{\alpha^n}{\sqrt{n!}}|n\ra.
\label{cohdef}\ee
In the coordinate representation these states are described by a trivial shifted form
of the vacuum wave functions
\beq
\psi_\alpha(x):=\la x|\alpha\ra=
e^{-\tfrac{1}{2}|\alpha|^2-\tfrac{1}{2}\alpha^2 +\sqrt{2}\alpha x -\tfrac{1}{2}x^2 }.
\label{cohstate}\ee

The Hamiltonian factorization \eqref{harmFact} plays a principally important role in the presented
formulas. However, as noticed in \cite{spi:nasa,S0}, this factorization is far from being unique.
In particular, one can define operator factors  in \eqref{harmFact} leading to the same Hamiltonian
in the following way:
$$
 A^+= \frac{-\textup{i}p+x}{\sqrt{2}}U(p,x), \qquad A=V(p,x)\frac{\textup{i}p+x}{\sqrt{2}},
 \qquad U(p,x) V(p,x)=1.
$$
If one demands that the operators $A$ and $A^+$ are hermitian conjugates of each other, then
$V(p,x)=U(p,x)^\dag$ and $U(p,x)U(p,x)^\dag =1$, i.e. $U(p,x)$ is a unitary operator.
In principle, $U$ may have an arbitrary admissible form, e.g. be a finite-difference or an
integral operator. The key question is -- what will be an explicit form of the operator
$AA^+=U(p,x)^\dag(L+\tfrac{1}{2})U(p,x)$ ? Let us demand that the Heisenberg algebra
\eqref{Heis} is preserved, i.e. that our tranformation is canonical, $[A, A^+]=1$.
This leads to the following equation
$$
U(p,x)^\dag(p^2+x^2)U(p,x)=p^2+x^2.
$$

Let us describe two simplest examples of nontrivial operators  $U(p,x)$, satisfying
this equality. First, this is the parity operator $P$,
\beq
U(p,x)=U(p,x)^\dag=P, \qquad Px=-xP,\qquad Pp=-pP, \qquad P^2=1,
\label{symmetry}\ee
which corresponds to the self-similar potential considered above for $N=1,\, q=-1, \, \mu=1.$
Second, this is the Fourier trans\-for\-ma\-tion operator  $U^\dag={\mathcal F}$,  $U ={\mathcal F}^{-1}$,
\beq
 [{\mathcal F}f](y):=\frac{1}{\sqrt{2\pi}}\int_{-\infty}^\infty e^{\textup{i} yx}f(x) dx,
 \qquad {\mathcal F}^2=P,\quad {\mathcal F}^4=1,
\label{Fourier}\ee
leading to the symplectic reflection,
\beq
{\mathcal F} x{\mathcal F}^{-1}=p, \qquad {\mathcal F} p {\mathcal F}^{-1}=  -x,
\label{FourierRefl}\ee
and preserving the canonical commutation relation $xp-px=\textrm{i}$.
The choice of the inverse Fourier trans\-for\-ma\-tion $U^\dag={\mathcal F}^{-1}$,
\beq
 [{\mathcal F}^{-1} f](y) =\frac{1}{\sqrt{2\pi}}\int_{-\infty}^\infty e^{-\textup{i} yx}f(x) dx,
\label{FourierInv}\ee
leads to another simplectic replection (a combination of the direct Fourier trans\-for\-ma\-tion
and parity)
$$
{\mathcal F}^{-1} x {\mathcal F} =- p, \qquad {\mathcal F}^{-1} p{\mathcal F} =x.
$$

Coherent states associated with the lowering operator $A=U^{-1} a$ are defined by the equation
\beq
A|\alpha\ra_U =\alpha|\alpha\ra_U, \quad  \text{or} \quad a|\alpha\ra_U =\alpha U|\alpha\ra_U.
\label{GCS}\ee
The choice $U=P$ was considered in \cite{spi:universal} and it leads to the following
 superposition of the canonical coherent states (``a Schr\"odinger cat'')
\beq
|\alpha\rangle_P=\frac{e^{- \pi\textup{i}/4}|\textup{i}\alpha\rangle
+e^{\pi\textup{i}/4}|-\textup{i}\alpha\rangle}{\sqrt2}.
\label{PCS}\ee
In the coordinate representation $\psi_\alpha^{(U)}(x):=\la x|\alpha\ra_U$ we have
\beq
\psi_\alpha^{(P)}(x)=
{\sqrt2\over \pi^{1/4}} \exp\left({\alpha^2-|\alpha|^2-x^2\over 2}\right)
\cos(\sqrt2 \alpha x - {\pi\over 4}).
\label{Pcoh}\ee
A detailed discussion of such superpositions of coherent states, as well as
a list of the correspinding literature sources is given in the paper \cite{spi:universal}.
We note that in this work a one-parameter generalization of the states \eqref{PCS}
is constructed which is also related to the parity operator through the choice
$U=\cos\varphi +\textup{i}P\sin \varphi$, but we shall not be considering it here.

The choice $U={\mathcal F}^{\pm 1}$ corresponds to the integro-differetial equations
\beq
\left(\frac{d}{dx}+x\right)\psi_\alpha^{({\mathcal F}^{\pm 1})}(x)=\frac{\alpha}{\pi^{1/2}}\int_{-\infty}^{\infty}
e^{\pm \textup{i}xy}\psi_\alpha^{({\mathcal F}^{\pm 1})}(y)dy, \quad x\in\R, \, \alpha\in \CC,
\label{FC1}\ee
which were not considered earlier in the literature. Normalizable solutions of these equations
are found with the help of the well known fact that eigenfunctions of the Fourier trans\-for\-ma\-tion
are given by the harmonic oscillator wave functions  \eqref{harmEigen}
\beq
L|n\ra=(n+\tfrac{1}{2})\, |n\ra,  \qquad
{\mathcal F}|n\ra=\varepsilon_n |n\ra,  \quad \varepsilon_n=e^{\frac{1}{2}\pi\textup{i}n}=\textup{i}^n.
\label{FourEigen}\ee

Let us expand $|\alpha\rangle_{{\mathcal F}^{\pm1}}$ in a series over the states $|n\ra$, which form
a basis of the Hilbert space,  $|\alpha\rangle_{{\mathcal F}^{\pm1}}=\sum_{n=0}^\infty c_n^{\pm}|n\ra$.
Substituting this expression  into the equation  \eqref{FC1}, we find recurrence relations of the
first order for the coefficients $c_n^{\pm}$, which are easily solved. As a result, we obtain
$$
|\alpha\rangle_{{\mathcal F}^{\pm1}}=e^{-\frac{1}{2}|\alpha|^2}\sum_{n=0}^\infty \frac{\alpha^n}{\sqrt{n!}}
e^{\pm \frac{1}{4}\pi\textup{i} n(n-1)}|n\ra, \quad
{}_{{\mathcal F}^{\pm1}}\la\alpha|\alpha\rangle_{{\mathcal F}^{\pm1}}=1.
$$
These wave functions also can be represented in the form of finite superpositions of the
canonical coherent states
\begin{eqnarray} \nonumber &&
|\alpha\rangle_{\mathcal F}=\tfrac{1}{2}\big( |e^{\frac{\pi \textup{i}}{4}}\alpha\ra
-e^{\frac{\pi \textup{i}}{4}}|e^{\frac{3\pi \textup{i}}{4}}\alpha\ra
+|e^{\frac{5\pi \textup{i}}{4}}\alpha\ra
+e^{\frac{\pi \textup{i}}{4}}|e^{\frac{7\pi \textup{i}}{4}}\alpha\ra\big),
\\ && \makebox[-2em]{}
\psi_\alpha^{({\mathcal F})}(x)=e^{-\frac{1}{2}(|\alpha|^2+x^2)}
\big(e^{-\frac{\textup{i}}{2}\alpha^2}\cosh ((1+\textup{i})\alpha x)
+e^{\frac{\textup{i}}{2}\alpha^2+\frac{\pi\textup{i}}{4}}
\sinh ((1-\textup{i})\alpha x) \big)
\label{FC1exp}\end{eqnarray}
and
\begin{eqnarray} \nonumber &&
|\alpha\rangle_{{\mathcal F}^{-1}}=\tfrac{1}{2}\big(
e^{-\frac{\pi \textup{i}}{4}}|e^{\frac{\pi \textup{i}}{4}}\alpha\ra
+|e^{\frac{3\pi \textup{i}}{4}}\alpha\ra
-e^{-\frac{\pi \textup{i}}{4}}|e^{\frac{5\pi \textup{i}}{4}}\alpha\ra
+|e^{\frac{7\pi \textup{i}}{4}}\alpha\ra\big),
\\ && \makebox[-2em]{}
\psi_\alpha^{({\mathcal F}^{-1})}(x)=e^{-\frac{1}{2}(|\alpha|^2+x^2)}
\big(e^{\frac{\textup{i}}{2}\alpha^2}\cosh ((1-\textup{i})\alpha x)
+e^{-\frac{\textup{i}}{2}\alpha^2-\frac{\pi\textup{i}}{4}}\sinh ((1+\textup{i})\alpha x) \big).
\label{FC2exp}\end{eqnarray}
Note that $\psi_\alpha^{({\mathcal F}^{-1})}(x)=(\psi_{\alpha^*}^{({\mathcal F})}(x))^*$,
where the star $^*$ denotes the complex conjugation.

Thus we have found all solutions of the integro-differential equations  \eqref{FC1}
belonging to $\text{L}^2(\R)$. It would be interesting to characterize other solutions
of these equations lying in more general functional spaces, or to find their
(and of the equation \eqref{GCSexp} below) general solution analogous to the
parabolic cylinder functions (or confluent hypergeometric functions)
characterizing general solution of the Schr\"odinger equation for the
harmonic oscillator potential.

Coherent states \eqref{PCS} describe superposition of the ``macroscopic'' states
exhibiting certain purely quantum mechanical experimental manifestations.
As far as the author knows, the ``Schr\"odinger cat'' states  \eqref{FC1exp} and \eqref{FC2exp},
related to the Fourier trans\-for\-ma\-tions, were not discussed in the literature.
Therefore it would be interesting to analyze their possible experimental manifestation.

The general symmetry trans\-for\-ma\-tion  $U(p,x)$ in \eqref{symmetry}, including into itself the
cases described above, is an arbitrary angle $\varphi\in\mathbb{R}$ rotation in the phase space,
\beq
U(\varphi)^\dag x U(\varphi) = x \cos \varphi+ p \sin \varphi, \qquad
U(\varphi) ^\dag p U(\varphi) =-x \sin \varphi+ p \cos \varphi.
\label{rotation}\ee
This is nothing else than the evolution in the artificial time $\varphi$
according to the Schr\"odinger equation
\beq
\textup{i}\partial_t\psi(t)=L\psi(t),\quad \psi(t)=U(t)\psi(0),\quad U(t)=e^{-\textup{i} tL}, \quad t=\varphi.
\label{TimeEvol}\ee
Validity of the trans\-for\-ma\-tions \eqref{rotation} is easy to prove at the level of
creation and annihilation operators. Indeed, since $U^\dag=U^{-1}=e^{\textup{i}tL}$,
\beq
a(\varphi):=e^{\textup{i}\varphi L}ae^{-\textup{i}\varphi L}, \quad
\partial_\varphi a(\varphi)=\textup{i}e^{\textup{i}\varphi L}[L,a]e^{-\textup{i}\varphi L}
=-\textup{i}a(\varphi).
\label{timerot}\ee
Solving this equation with the initial condition $a(0):=a$,
we find $a(\varphi)=e^{-\textup{i}\varphi}a$. Analogously,  $a^+(\varphi)=e^{\textup{i}\varphi L}a^+e^{-\textup{i}\varphi L}=e^{\textup{i}\varphi}a^+$, which in combination yields \eqref{rotation}.

For this general case equation \eqref{GCS} determining coherent states takes the maximally
complicated integro-differential form:
\beq
\left(\frac{d}{dx}+x\right)\psi_\alpha^{(U)}(x)= e^{-\textup{i} \varphi L}\psi_\alpha^{(U)}(x)
=\sqrt{2}\alpha\int_{-\infty}^{\infty} \mathcal{K}(x,y;\varphi)\psi_\alpha^{(U)}(y)dy,
\label{GCSexp}\ee
where the kernel of the integral operator on the right-hand side has the following explicit form
\beq
\mathcal{K}(x,y;\varphi)=\frac{1}{\sqrt{2\pi\textup{i}\sin \varphi}}
\exp\big[ \textup{i}\frac{(x^2+y^2)\cos \varphi -2xy}{2\sin \varphi}\big].
\label{Mehler}\ee
This kernel was constructed by Mehler 60 years before the discovery of quantum mechanics
in terms of the variable corresponding to the replacement $\varphi\to -\textup{i}\varphi$.
The integral trans\-for\-ma\-tion itself standing on the right-hand side of the equality \eqref{GCSexp},
can be interpreted as a ``fractional'' generalization of the standard Fourier trans\-for\-ma\-tion,
which corresponds to the parameter choice $\varphi=\pm \pi/2$ (see, for instance \cite{SS}).

The normalizable solution of equation \eqref{GCSexp} is uniquely found by the expansion over
the basis of eigenfunctions of the operator $L$ and solving the corresponding recurrence relation
 for the expansion coefficients. This yields the expression
\beq
|\alpha\rangle_U=e^{-\frac{1}{2}|\alpha|^2}
\sum_{n=0}^\infty \frac{\alpha^n}{\sqrt{n!}} e^{-\textup{i} \varphi \frac{n(n-1)}{2}}|n\ra,
\qquad U={e^{-\textup{i} \varphi L}}.
\label{TG}\ee
For arbitrary values of the parameter $\varphi$ these wave functions cannot be written anymore
as finite combinations of the canonical coherent states -- they represent an example of what is
called the Titulaer-Glauber coherent state  which were discussed in \cite{spi:universal}.
We stress that here  $\varphi$ is not the real time in the Schr\"odinger equation,
but an ordinary parameter characterizing the wave function \eqref{TG}. If we consider
the real time evolution then it looks as follows
$$
|\alpha,t\rangle_U:=e^{-\textup{i}t L}|\alpha\rangle_U=e^{-\frac{1}{2}|\alpha|^2}
\sum_{n=0}^\infty \frac{\alpha^n}{\sqrt{n!}} e^{-\textup{i} \varphi \frac{n(n-1)}{2}}
e^{-\textup{i}t (n+\frac{1}{2})}|n\ra=e^{-\textup{i}\frac{t}{2}}|e^{-\textup{i}t}\alpha\rangle_U.
$$

Now let us shortly consider coherent states of the general class of self-similar
potentials.  Suppose that  $0<q^2<1$. Then for the negative values of the energy, $\lambda<0$,
the lowering operator is $B$. The corresponding coherent  states are defined in the standard way
\cite{spi:universal}
\beq
B|\alpha\rangle^{(k)}_-=\alpha |\alpha\rangle^{(k)}_-,\qquad \la x |\alpha\rangle^{(k)}_-\propto {}_0\varphi_{N-1}(\ldots;q^2,z)\psi_k(x),
\label{B-}\ee
where $\psi_k(x),\, k=0,\ldots, N-1$, are the first $N$ eigenfunctions of $L$,
and the operator argument $z$ is proportional to the raising operator $B^+$.
These states are described by the standard basic hypergeometric series
${}_0\varphi_{N-1}$ \cite{aar:special}.

For $\lambda>0$ the lowering operator is $B^+$, which leads to the coherent states of the
principally new type \cite{spi:universal}
\beq
B^+|\alpha\rangle^{(s)}_+=\alpha |\alpha\rangle^{(s)}_+, \quad s\in\mathbb{Z}.
\label{B+}\ee
These states are described  by the bilateral  basic hypergeometric series
${}_0\psi_{N-1}$ or the Ramanujan integrals.

Let $N=1$, $u(x)=0$, $0 < q < 1$ and
$$
U=q^{1/2+x\partial_x}, \qquad U^\dag=U^{-1}=q^{-1/2-x\partial_x}.
$$
Then the operators
\beq
A=q^{-x\partial_x-1/2}\Big(\partial_x+\frac{1}{\sqrt{1-q^2}}\Big),
\quad A^+=\Big(-\partial_x+\frac{1}{\sqrt{1-q^2}}\Big)q^{x\partial_x+1/2}.
\label{freeqalg}\ee
satisfy the $q$-oscillator algebra $AA^+-q^2A^+A=1$.
For this trivial system -- the free nonrelativistic quantum particle, there exist
normalizable coherent states  $A^+\psi_\alpha^+(x)=\alpha\psi_\alpha^+(x)$,
$\psi_\alpha^+(x)\in \mathrm{L}^2(\mathbb{R})$.  They are determined as solutions of
the advanced pantograph equation  \cite{S2}
\beq
\partial_x\psi_\alpha^+(x)=-\alpha q^{-3/2}\psi_\alpha^+(q^{-1}x)
+\frac{q^{-1}}{\sqrt{1-q^2}}\psi_\alpha^+(x),
\label{pantograph}\ee
admitting an infinite number of solutions lying in   $\text{L}^2(\R)$.
Discussion of the explicit form of these coherent states and some of their
properties is given in   \cite{spi:universal}. As far as the author knows,
possible physical manifestation of these states in quantum optics was not discussed yet.

\section{Solitons}

Consider now evolution of the potentials in a different continuous time
$u_j(x)\to u_j(x,t)$, which is not related to the time evolution in the Schr\"odinger
equation and determined by the following differential law
\beq
\partial_t\psi^{(j)}(x,t)=D_j\psi^{(j)}(x,t), \quad
D_j:= -4\partial_x^3+6u_j(x,t)\partial_x+3\partial_xu_{j}(x,t).
\label{LA}\ee
From the quantum mechanics point of view here $t$ is simply a potential parameter.
Compatibility of \eqref{LA} with the Schr\"odinger equation leads to the operator relation
$\partial_t L_j=[D_j, L_j]$, which is equivalent to the Korteweg-de Vries (KdV) equation \cite{sol:book}
\beq
\partial_tu_j(x,t)-6u_j(x,t) \partial_x u_{j}(x,t)+\partial_x^3 u_{j}(x,t)=0.
\label{KdV}\ee
Note that evolution law \eqref{LA} can be obtained with the help of limiting relation
in the discrete time $j$ in the Infeld chain. For this it is suffiicient to consider
the evolution in $j$ three steps forward $j\to j+3$, define $t=jh$ and consider
the limit   $h\to 0$ for fixed $t$ in the evolution law
$\psi^{(t/h+3)}=M^{(3)}_{t/h}\psi^{(t/h)}$. For special choice of the differential
operator of the third order $M^{(3)}_{t/h}$ there emerges the relation \eqref{LA}.

Let us substitute the ansatz $f_j = -\partial_x \log \phi_{0}^{(j)}$ in the definition
of potentials $u_j=f_j^2-f_{j}'+\lambda_j$. This results in a linear differential equation
\beq
-\partial_x^2\phi_{0}^{(j)}+u_j\phi_0^{(j)}=\lambda_j\phi_0^{(j)},
\label{phi0}\ee
i.e. $\phi_0^{(j)}$ is an eigenfunction of $L_j$ for $\lambda=\lambda_j$ and
\beq
\psi^{(j+1)} = \frac{\phi_0^{(j)}\partial_x \psi^{(j)}  -\psi^{(j)}\partial_x \phi_{0}^{(j)}}{\phi_0^{(j)}}
=\frac{W(\phi_0^{(j)}, \psi^{(j)})}{\phi_0^{(j)}}.
\label{darboux}\ee

Let $\phi_k^{(j)}$, $k=0,\ldots,n-1$, be formal eigenfunctions of the operator  $L_j$
with the eigenvalues $\lambda_{j+k}$, i.e. $L_j\phi_k^{(j)}=\lambda_{j+k}\phi_k^{(j)}$.
Then according to the paper \cite{Crum}
\beq
u_{j+n}(x,t)=u_j(x,t)-2\partial_x^2\log W(\phi_0^{(j)},\dots,\phi_{n-1}^{(j)}),
\label{Dar}\ee
where $W(\phi_0,\dots,\phi_{n-1})=\det(\partial_x^i\phi_k)$ is the Wronskian of $n$ functions $\phi_k(x)$.
Analogously, one can   write explicitly
\begin{eqnarray}\label{fn} &&
f_{j+n}(x,t)=-\partial_x\log \frac{W\left(\phi_0^{(j)},\dots,
\phi_n^{(j)}\right)}{W\left(\phi_0^{(j)},\dots,\phi_{n-1}^{(j)}\right)},
\\ &&
\psi^{(j+n+1)}(x,t)=\frac{W\left(\phi_0^{(j)},\dots,\phi_n^{(j)},\psi^{(j)}\right)}
{W\left(\phi_0^{(j)},\dots,\phi_n^{(j)}\right)}=M_j^{(n+1)} \psi^{(j)}(x,t).
\label{Crum}\end{eqnarray}

The variable $t$ does not play any role in the consideration of the Darboux trans\-for\-ma\-tion
\eqref{Dar}, it is completely independent of the discrete time used in the
Infeld factorization chain. Therefore, if  $u_j(x,t)$ satisfies the KdV-equation, then
$u_{j+n}(x,t)$ also will be a solution of this equation. Correspondingly, starting from a
simple solution of this equation one can build more and more complicated solutions.

The most refined description of the integrable systems is reached in terms of
the so-called tau-function, which already appeared earlier in an implicit form.
It is defined in the following way
\beq
u_j(x,t)=-2\partial_x^2 \log\tau_j(x,t).
\label{tau}\ee
This function is quite convenient because its zeros determine double poles of the potential in  $x$.
If $\tau_j(x,t)$ is a holomorphic function of its arguments $x$ and $t$, as it was in the case of
finite-gap potentials \eqref{finitegap}, then potentials are meromorphic functions on the whole
complex plane, which is a strong restriction on the class of considered potentials.
Relations \eqref{fn} and (\ref{Crum}) take a substantially more compact form in terms of the
$\tau$-function
\beq
\tau_{j+n}=W(\phi_0^{(j)},\dots,\phi_{n-1}^{(j)})\tau_j,
\qquad f_j=-\partial_x\log \tau_{j+1}/\tau_j.
\label{tautrans}\ee

In terms of another dependent  variable one can write
\beq
\rho_j:=-\partial_x\log\tau_j,\qquad u_j=2\partial_x\rho_{j},\quad
f_j=\rho_{j+1}-\rho_j.
\label{rhodef}\ee
As a result of such a change of variables the relation between $u_j(x)$ and $f_j(x)$
takes onto iself the role of the factorization chain
\beq
\frac{d}{dx}(\rho_{j+1}+\rho_j)-(\rho_{j+1}-\rho_j)^2=\lambda_j.
\label{rhochain}\ee
We remark that this equation has a simple shifting symmetry
$\rho_j(x)\to \rho_j(x)+cx+const$, $\lambda_j\to \lambda_j+2c$.

Self-similar potentials are characterized by very simple restrictions
on the $\tau$-function, $\tau_{j+N}(x,t)=\tau_j(qx,q^3 t)$, or
$\rho_{j+N}(x,t)=q\rho_j(qx,q^3 t)$, and the spectral constants $\lambda_{j+N}=q^2\lambda_j$.
For $N=1$ this leads to the mixed differential and $q$-difference nonlinear equation of the form
\beq
\frac{d\rho(x)}{dx}+q\frac{d\rho(qx)}{dx}- (q\rho(qx)-\rho(x))^2=\mu,
\label{rhoselfsim}\ee
where $\rho(x):=\rho_0(x)$ and $\mu:=\lambda_0$.

For $u_0(x,t)=0$, let us take $\lambda_j=-k_j^2/4$, $\theta_j:=k_jx-k_j^3t+\theta_j^{(0)}$,
$k_j,\, \theta_j^{(0)}\in\R$, and
\beq
\phi_{2j}^{(0)}=\cosh \tfrac{1}{2}\theta_{2j}(x,t),\quad
\phi_{2j+1}^{(0)}=\sinh \tfrac{1}{2}\theta_{2j+1}(x,t), \; j=0,1,\ldots, n-1.
\label{solit}\ee
This choice results in  $u_n(x,t)$ which is a nonsingular reflectionless potential called
the $n$-soliton solution of KdV-equation, since it describes solitary waves on the
shallow water \cite{MS}. In this picture $k_j^2$ is proportional to the amplitudes and speeds of
the solitons, and $\theta^{(0)}_j$ describe soliton phases (i.e. their mutual position at $t=0$).
Beyond the given Wronskian representation, there exist other explicit expressions for the
$n$-soliton solutions of nonlinear integrable equations. For instance, for the KdV-equation
one can write
\begin{eqnarray} 
u_n(x,t)=-2\partial_x^2 \log \tau_n(x,t), \quad
\tau_n(x,t)=\det C,
\quad C_{ij}=\delta_{ij}+\frac{2\sqrt{k_ik_j}}{k_i+k_j} e^{(\theta_i+\theta_j)/2}.
\label{nsol}\end{eqnarray}
In the next section we give another representation having a bright physical interpretation.

\section{Ising chains and the lattice Coulomb gas}

In 1971 Hirota \cite{Hir} has found the following explicit representation of the $n$-soliton tau-function \eqref{nsol}
\beq
\tau_n = \sum_{\sigma_i=0,1} \exp \left( \sum_{0\leq i<j \le n-1 } A_{ij}
\sigma_i \sigma_j + \sum_{i=0}^{n-1} \theta_i\sigma_i \right), \quad
e^{A_{ij}}={ (k_i-k_j)^2 \over  (k_i+k_j)^2 },
\label{Hirota}\ee
where the variables $A_{ij}$ describe the scattering phases of solitons.
For experts in statistical mechanics it is not difficult to recognize in the expression
\eqref{Hirota} the partition function of a one-dimensional lattice gas.
In this picture the variable $\sigma_i$ is the filling number of the $i$-th cell.
For $\sigma_i=0$ the cell is free and for $\sigma_i=1$ the cell is occupied.
At the same time, the coefficents  $A_{ij}$ are proportional to the interaction potential
between $i$-th and $j$-th molecules, and  $\theta_i$
serve as local chemical potentials. Inspite of the evidence of such an interpretation,
this relation between solitonic solutions of the nonlinear integrable equations and
partition functions was discovered only in 1997 in the paper \cite{lou-spi:self}
(see also \cite{lou-spi:spectral}).

Let us change the variables  $s_i=2\sigma_i-1=\pm 1$ in the expression \eqref{Hirota}.
This brings us to an one-dimensional  Ising model with a nonlocal exchange:
\begin{eqnarray}\nonumber && \makebox[3em]{}
\tau_n= e^{\varphi}Z_n, \qquad \varphi=\frac{1}{4}\sum_{i<j} A_{ij}
+ \frac{1}{2}\sum_{j=0}^{n-1} \theta_j,
\\  &&
Z_n=\sum_{s_i=\pm1} e^{- \beta E}, \qquad
E =\sum_{0\leq i<j \leq n-1} J_{ij}s_is_j -\sum_{0\leq i=\leq n-1} H_i s_i,
\nonumber \\ &&
\beta J_{ij} =-\frac{1}{4} \; A_{ij},
\qquad
\beta H_i=\frac{1}{2}\theta_i +\frac{1}{4}\sum_{0\leq j\neq i\leq n-1} A_{ij},
\quad \beta=\frac{1}{kT}.
\label{Ising}\end{eqnarray}
In this picture the variable  $s_i$ describes a spin located at the $i$-th cell,
$J_{ij}$ are the exchange constants, and $H_i$ is an inhomogeneous
(i.e. depending on $i$) external magnetic field. At the same time from the explicit form of
intermolecular potential $\propto A_{ij}$ or exchange constants  $J_{ij}$ it follows
that we have a substantial restriction on the given interpretation of the soliton tau-function
as grand canonical partition function of a lattice gas or partition function
of a nonlocal Ising model in the external magnetic field. Namely, the temperature of
the system is fixed.

Let us consider now a self-similar infinite soliton system, i.e. the limit
$n\to\infty$ under the following self-similar restriction on the soliton parameters
\beq
\theta_{j+N}(x,t)=\theta_j(qx,q^3t),  \quad \text{or} \quad
k_{j+N}=qk_j, \qquad \theta_{j+N}^{(0)}=\theta_j^{(0)}.
\label{selfsimsol}\ee
Under such constraints we have $\tau_{j+N}(x,t)=\tau_j(qx,q^3t)$.
Take the limiting infinite soliton potential $u_\infty(x,t)$.
It exhibits such an evident property  that the scaling trans\-for\-ma\-tion of the
coordinate  $x\to qx$ and time $t\to q^3 t$ induces the potential trans\-for\-ma\-tion
$u_\infty(x,t)\to q^2u_\infty(qx,q^3t)$, which deletes  $N$ solitons
corresponding to the lowest  $N$ eigenvalues of the Hamiltonian.

From the described above Ising model point of view, self-similarity of the
Hamiltonian spectrum is equivalent to the translational invariance of the
exchange constants for spins  $J_{i+N,j+N} = J_{ij}$ and external magnetic field $H_{j+N}=H_i$.
This constraint makes it possibile to compute exactly the free energy per cell in the
thermodynamic limit, i.e. the asymptotics of the partition function  $Z_n$  for $n\to\infty$.
In particular, for the minimal period $N=1$, leading to the homogeneous magnetic field $H_i=H$,
 there emerges the following expression \cite{lou-spi:self}:
\begin{eqnarray}\nonumber  &&
m(H)= \partial_{\beta H}\log Z_n =
\stackreb{\lim}{n\to\infty} n^{-1}\sum_{i=0}^{n-1}\langle s_i \rangle
=\stackreb{\lim}{n\to\infty} n^{-1}\partial_{\beta H}\log Z_n
\\   && \makebox[2em]{}
=\left(1-\frac{1}{\pi}
\int_0^{\pi}\frac{\theta_1^2(\nu, q^2)d\nu}{\theta_4^2(\nu, q^2)\cosh^2\beta H
-\theta_1^2(\nu, q^2) \sinh^2\beta H} \right)\tanh \beta H,
\label{magnetization}\end{eqnarray}
where $\theta_{1,4}(\nu, q^2)$ are the Jacobi theta functions.
Analogous situation holds true for soliton solutions of the Kadomtsev-Petviashvili (KP)
equation and for a number of other integrable equations. Despite the one-dimensionality of
the model, because of the nonlocality of interaction there emerges in the thermodynamic limit
a nontrivial phase transition corresponding to the effective temperature
$T=0$ induced by the limit $q\to1$ \cite{lou-spi:critical}.

The most general picture related to the described physical interpretation consists in a connection with
the two-dimensional Coulomb gas on the plane with different boundary conditions. Indeed,
consider the grand canonical partition function of
$n$ charged particles on the plane which are allowed to sit only in
a discrete set of points on some lattice $\Gamma$,
\begin{equation}\label{Coulomb}
Z_n=\sum_{\sigma(z_i)=0,1}\exp\Bigl(\frac{1}{2}
\sum_{i\neq j}W(z_i,z_j)\sigma(z_i)\sigma(z_j)
+\sum_{z_i\in\Gamma}\theta(z_i)\sigma(z_i)\Bigr),
\end{equation}
where $z_j=x_j+\textup{i}y_j$ are the coordinates of the lattice $\Gamma$ vertices,
$\sigma(z_i)=1$, if the point $z_i$ is occupied by the Coulomb particle with the charge
$q(z_i)$, and $\sigma(z_i)=0$, if this point on the lattice is free. As a result
\begin{equation}\label{CoulmbEnergy}
W\left(z,z^\prime\right)=-\beta E_n, \quad E_n= q(z)q(z^\prime)V(z,z^\prime),
 \quad  V(z,z^\prime)=-\ln |z-z^\prime|,
\ee
where $E_n$ is the interaction energy of $n$ Coulomb particles on the plane. Moreover,
\begin{equation}\label{CoulombField}
\theta(z)=\mu(z)-\beta\left(q(z)v(z)+q(z)\phi(z)\right),
\ee
where $v(z)$ describes the interaction of charges with their artificial
images appearing from the boundary conditions (the conducting boundary surface
or the dielectric), $\phi(z)$ is the external electric field, and $\mu(z)$
is the local chemical potential.

By a special choice of the lattice $\Gamma$, of the type of boundary conditions
and external fields, connection between the complex spectral variables  $k_i$
and the coordinates $z_i$  one can reproduce the $n$-soliton tau-functions, $Z_n=\tau_n$,
of different integrable equations, including KdV, KP, the Toda chain and so on.
The detailed description of such a relation is given in the paper \cite{lou-spi:soliton}.

There are other important applications of solitons showing their versatile nature.
For example, corresponding potentials provide partial solution of the Hadamard problem
of constructing the wave operators satisfying the Huygens' principle  \cite{BL}.
Their degenerate form describes solution of the electrostatics problems for particles
of different charges on the plane \cite{L1,L2}). A two-dimensional setup of the factorization
method is useful for building exact solutions in some reduced problems of fluid dynamics
(the Hele-Shaw problem with varying coefficients) \cite{L3}. Let us mention also that the
solitonic two-dimensional Coulomb gas systems described above are related to the Laplacian
growth \cite{L4}.

\section{Discrete Schr\"odinger equation}

Evolution in time \eqref{DTE} breaks the normalization of wave functions, i.e. if the
initial  wave functions are normalized to unity, then the shift in time breaks this property.
For removing this drawback it is necessary  to renormalize the evolution law
\beq
\psi^{(j+1)}(x)=\frac{A_j}{\sqrt{\lambda-\lambda_j}}\psi^{(j)}(x), \qquad
\psi^{(j)}(x)=\frac{A_j^+}{\sqrt{\lambda-\lambda_j}}\psi^{(j+1)}(x).
\label{DTE2}\ee
Now it is easy to check that normalizations of the Hamiltonian  eigenfunctions do not change
$$
\int_{-\infty}^\infty|\psi^{(j+N)}(x)|^2dx=\int_{-\infty}^\infty|\psi^{(j)}(x)|^2dx=1,
$$
under the condition that zero modes of the evolution operators are not considered.
Taking two steps evolution in time
$$
\psi^{(j+1)}(x)=\frac{A_j}{\sqrt{\lambda-\lambda_j}}\frac{A_{j-1}}{\sqrt{\lambda-\lambda_{j-1}}}\psi^{(j-1)}(x)
$$
and removing derivatives of wave functions on the right-hand side either
with the help of the Schr\"odinger equation, or with the help of relations \eqref{DTE2},
we obtain a three term recurrence relation
\beq
\sqrt{\lambda-\lambda_{j}}\psi^{(j+1)}(x)-(f_j(x)+f_{j-1}(x))\psi^{(j)}(x)
+\sqrt{\lambda-\lambda_{j-1}}\psi^{(j-1)}(x) =0,
\label{TTR}\ee
where the coordinate $x$ enters as a fixed parameter and $\lambda$ remains the spectral parameter
as before. Thus for an arbitrary initial potential $u_0(x)$ solutions of the Infeld factorization
chain with an analytical dependence on the discrete time $j$ lead to the finite-difference
equation of the second order  $j$, which in turn  can be considered as a finite-difference
analogue of the Schr\"odinger equation.

The harmonic oscillator gives the simplest example  when the discrete spectrum wave functions
are described by the orthogonal polynomials. All orthogonal polynomials satisfy
a three term recurrence relation which can be represented in the form   \cite{Szego}
\beq
p_{n+1}(x)+u_np_{n-1}(x)+b_np_n(x)=xp_n(x), \quad n\geq 0, \quad p_{-1}=0, \quad p_0=1,
\label{TTRpol}\ee
generating monic polynomials $p_n(x)=x^n+\ldots$. The orthogonality measure will be
positively defined, if the recurrence coefficients take finite values and
satisfy the constraints $u_n, b_n\in\R$, $u_n>0$.

If one abandons the boundary conditions in  \eqref{TTRpol} and, correspondingly,
the polynomiality of the eigenfunctions $p_n(x),\, n\in\Z$, there emerges
a discrete Schr\"odinger equation on the lattice of integer numbers.
An application of the factorization method to difference equations of the second order
on the basis of ``old'' orthogonal polynomials was considered for the first time in
the paper  \cite{Mi1}. However, this was done in the spirit of Schr\"odinger, i.e.
with the help of concrete known solvable equations. An approach in the spirit of Infeld was
considered substantialy later from the viewpoint of the Toda chain and other discrete
integrable systems.

Let us consider the following difference equations in two
discrete variables $n$ and $j$
\begin{eqnarray} \label {CT} &&
p_n^{j+1}(x)=\frac{p_{n+1}^{j}(x)+C_n^{j+1}p_n^{j}(x)}{x-\lambda_{j+1}},
\\  \label{GT} &&
p_n^{j-1}(x)=p_n^{j}(x)+A_n^{j}p_{n-1}^{j}(x),
\end{eqnarray}
where $A_n^{j}, \, C_n^{j+1}$ are some indetermined coefficients.
These equations serve as discrete analogues of the evolution laws for
the usual Schr\"odinger equation  \eqref{evolution} under the shifts  $j\to j\pm 1$.
The compatibility condition of the laws \eqref{CT} and
\eqref{GT} leads to the three term recurrence relation   \eqref{TTRpol},
where all variables except of $x$ should acquire the same upper index  $j$.
At the same time the following relation between recurrence coefficients is established
\beq
u_n^j=A_n^j C_n^j,\qquad b_n^j=A_{n+1}^j+C_n^j+\lambda_j.
\label{TTRfact}\ee
An analogue of the Infeld factorization chain has the form of a systems of two equations
\begin{eqnarray} \nonumber &&
A_n^{j+1}C_{n-1}^{j+1}=A_n^{j}C_n^{j},
 \\  &&
A_n^{j+1}+C_n^{j+1}+\lambda_{j+1}=A_{n+1}^{j}+C_n^{j}+\lambda_j,
 \label{DTTL2}\end{eqnarray}
which is called the discrete time Toda lattice. One can reformulate
the relations given above on the operator language, when $L_j$ is not a
Schr\"odinger operator, but a tridiagonal Jacobi matrix. Then the equations
\eqref{TTRfact} and \eqref{DTTL2} will be equivalent to the same operator relations
\eqref{fact}, but we omit this description here.

In the theory of orthogonal polynomials framework relations \eqref{CT}
are called the Christoffel spectral trans\-for\-ma\-tions, since they generate Christoffel's
kernel polynomials (i.e. they map polynomials to polynomials) \cite{Szego}.
Relations \eqref{GT} are called the Geronimus trans\-for\-ma\-tions \cite{Ger2}
and under certain circumstances they become inverses to the Christoffel trans\-for\-ma\-tions.
Often all these trans\-for\-ma\-tions are called discrete Darboux trans\-for\-ma\-tions,
although historically it is the Darboux trans\-for\-ma\-tions should be called the
continuous Christoffel trans\-for\-ma\-tions (which is reflected in the title
of Krein's paper \cite{Krein}). Note also that the isospectral version
$\lambda_j=const$ of equations \eqref{DTTL2} was constructed long before the
surge of interest to the theory of integrable systems within
numerical methods of the applied analysis \cite{Rutis}.

A discrete analogue of the ansatz \eqref{SepVar} for Infeld's factorization chain
was constructed in the papers \cite{SZ0,SZ1}. We shall not describe it here,
as well as discrete analogues of the self-similar potentials  \eqref{reduction},
since it requires many additional explanations. Let us only remark that
this ansatz reproduces the most general system of classical orthogonal polynomials constructed
by Askey and Wilson, and additionally it generates another system of orthogonal polynomials.
A systematic consideration of spectral trans\-for\-ma\-tions for orthogonal polynomials
on the basis of the Stieltjes function is given in \cite{alex}.

In the paper \cite{spi-zhe:spectral} the most general factorization chain is constructed
which is related to the polynomial systems or, equivalently, to biorthogonal
rational functions. In the same work there was formulated an ansatz of the
generalized separation of variables which had led  to a principally new family of
biorthogonal functions expressed in terms of the elliptic hypergeometric functions \cite{spi:essays}.
A description of this new class of special functions of mathematical physics,
which has found very important applications in the quantum field theory \cite{Spepan},
also goes beyong the present survey scope.

\section{Self-similarity and special functions}

Exactly solvable models of physical phenomena play an important methodological role.
They allow one to determine the domain of applicability of the models themselves
as well as to justify by rigorous mathematical methods their predictions.
Extension of the set of such examples is a central problem of mathematical physics.
It is necessary to explain for clarity the meaning of the term ``exact solution''
-- is it possible to give to it mathematical  formulation which does not
assume any links to the intuitive understandings? Here the key objects are the
elementary functions and their generalizations known as the special functions.
Everything is clear with the elementary functions -- these are the combinations
(mathematically called the ``fields'') of rational, power, exponential
and trigonometric functions and their inverses (i.e. radicals, logarithms, etc).
But there does not exist a universal definition of the special functions.
From the practical point of view -- these are the functions given in the handbooks
of special functions. However, one needs a characterization of their common properties
which would allow a constructive search of new such functions deserving a
proper place in the handbooks.

There are many handbooks and textbooks on the special functions, for example
\cite{aar:special,erd:higher,grad}. However, even the latest project of such scale
finished after ten years of work  \cite{DLMF} does not cover such well known functions
as the Painlev\'e transcendents or the elliptic hypergeometric functions \cite{spi:essays}.
Moreover, none of these books contains a list of formal requirements that a function
should satisfy in order to be called ``special''. Usually one discusses classes of
functions of specific form or properties, such as the hypergeometric,
elliptic, modular functions, and so on. For special functions it is natural to expect
that known local behavior of a function should admit a computation of the asymptotic of the
function at infinity, i.e. the asymptotic connection problem
should be solvable. Such an approach to special functions is characteristic to
investigations of the Painlev\'e type functions and the general theory of
isomonodromic deformations  \cite{kit}.

The group theory and related to it algebras provide a sufficiently rich set
of tools for building special functions, but historically their representations
theory yielded mainly interpretations of already known functions \cite{vil}.
Nevertheless, the general approach based on the groups of symmetries is
central for the theory of special functions. In particular, special functions
of the XIX century  emerged from the separation of variables in very
simple (and, thus, useful and universal) partial differential equations.
And at the basis of separation of variables one has the symmetries of those
equations. In the framework of our approach to special functions
they emerge as a result of self-similar reductions of
infinite chains  of spectral tansformations for linear eigenvalue problems.

This definition interprets special functions as the object tied to the fixed
points of different continuous and discrete trans\-for\-ma\-tions mapping the space of
solutions of a taken spectral problem onto itself. It is known that such an approach works
well in the case of functions of one independent variable, but even for them it does
not pretend on the coverage of all possible cases.  Note that these functions still may
depend on the infinite number of parameters. On the one hand, this definition is
tied to the theory of completely integrable systems \cite{sol:book}, for which
searches of self-similar solutions of the nonlinear evolution equations is
the standard problem. On the other hand, from the point of view of special functions
themselves, this approach is based on the contiguous relations -- linear or nonlinear
relations connecting special functions at different values of their parameters.

Let us summarize this half-heuristic scheme of building special functions.
One takes, as a germ, some linear spectral problem determined by a differential,
finite-difference, or integral equation (the author did not work with the
latter type of spectral problems). On the full space of solutions of this equation
one builds other linear equations in the variables entering as parameters.
That is one searches nontrivial operators under the actions of which the space of
solutions of the initial equation is mapped onto itself.

The compatibility condition of the taken system of linear equations leads to
nonlinear relations for the functions entering as free coefficients.
If both equations are differential, then one get the equations
of KdV, KP type and so on. One can get the mixed differential-difference cases
of the type of Infeld or Toda chains. The equations analogous to the Toda chain
with the discrete time \eqref{DTTL2} correspond to the completely finite-difference
schemes changing the spectral data  of the initial spectral problem in a prescribed
way. After that one performs an analysis
of the discrete and continuous symmetries of the derived nonlinear equations with the
help of Lie group-theoretical methods \cite{Yamilov}, which map the space of solutions onto itself.
At the final step one constructs self-similar solutions of the derived nonlinear equations
 which are invariant under some taken symmetries. In a result of such reductions there
 emerge closed systems of nonlinear differential, differential-difference,
 two-dimensional difference, etc equations, whose solutions define the ``nonlinear'' special
 functions (e.g., the described above continuous    $q$-analogues of the Painlev\'e functions).
 Solutions of the initial linear equations with the coefficients defined by the indicated
 self-similar functions determine ``linear'' special functions (e.g. functions of
 the hypergeometric  type). The latter two steps require applications of the heuristic
 thoughts, becasue there are no completely regular  ways of solving the corresponding problems.
 For instance, the ansatzes of the generalized separation of variables, used in \cite{SZ0,SZ1}
  for building recurrence relations for the associated Askey-Wilson polynomials and
  in  \cite{spi-zhe:spectral} in the discovery of elliptic biorthogoanl rational functions
  did not get a regular group-theoretical description yet.

Another important constituent element of this scheme of building special functions is the
transcendency theory. It is known that the Painlev\'e functions are transcendental over
the differential fields constructed by a finite number of Picard-Vessiot extensions over
the field of rational functions. Analogously, for a given solution of a taken equation
it is necessary to clarify which differential or finite-difference field it belongs to,
in particular, to answer the question whether it belongs to the field of coefficients of
this equation. So, the problem of interpretation of self-similar solutions of the Infeld
factorization chain from the point of view of differential or difference Galois
theory is open until now.

\section{Conclusion}

We have described several important physical applications of the functions
emerging from self-similar reductions of the Infeld factorization chain --
to solvable problems in quantum mechanics, to coherent states, solitons,
Ising chains and two-dimensional Coulomb gases. Additionally, we have presented
a number of graceful mathematical constructions in the context of the theory
of special functions, $q$-deformed algebras and unusual differential-difference equations.
The wide scope of applications of self-similar systems and a general interest to them
had led V. B. Priezzhev and the author to an idea of organizing big conference dedicated
to the corresponding thematics. This conference took place during two weeks at BLTP JINR
in the summer of 1998 and its results are reflected in the proceedings \cite{PS}.
I am deeply indebted to Vyacheslav Borisovich for a sincere interest to my research
and general intellectual support during all the time we knew each other.

The list of literature sources presented below does not pretend on the completeness.
There are many other surveys of the intersecting subjects, in particular,
\cite{BS,MR,AI}. One of the interesting subjects skipped here consists in a beautiful
interpretation of the factorization method in the framework of the supersymmetry concept.
Personally for me this application played a crucial role in the change of
the subject of my investigations, which  started from the work  \cite{RS}.
In particular, a development of the corresponding ideas had led to an interpretation of
the polynomials relations  \eqref{poly} as a nonlinear realization of the
supersymmetry algebra  \cite{AIS}.  A detailed review of such a generalization of
the supersymmetric quantum mechanics is given in  \cite{AI}.

\end{document}